# SoundScape: A Human-AI Co-Creation System Making Your Memories Heard

Chongjun Zhong

Zhejiang University, zhong_chongjun@zju.edu.cn

Jiaxing Yu

Zhejiang University, yujx@zju.edu.cn

Yingping Cao

Zhejiang University, caoyingping@zju.edu.cn

Songruoyao Wu

Zhejiang University, wsry@zju.edu.cn

Wenqi Wu

Zhejiang University, wenqi_wu@zju.edu.cn

Kejun Zhang

Zhejiang University, zhangkejun@zju.edu.cn

Sound plays a significant role in human memory, yet it is often overlooked by mainstream life-recording methods. Most current UGC (User-Generated Content) creation tools emphasize visual content while lacking user-friendly sound design features. This paper introduces SoundScape, a human-AI co-creation system that allows users to easily create sound memories on mobile devices through innovative interaction. By integrating sound effects and music with visual scenes, SoundScape encourages users to enrich their creations with immersive sound elements, enhancing the atmosphere of their works. To support public creation, SoundScape incorporates a conversational agent and AI music generation technology. User studies indicate that our approach is effective for sound memory creation, with SoundScape outperforming existing tools in terms of user experience and the perceived quality of produced works.

CCS CONCEPTS • **Human-centered computing Human computer interaction (HCI)~Interactive systems and tools** • **Applied computing~Arts and humanities~Sound and music computing**

## 1 INTRODUCTION

Memory is not just a record of our past, but also an important medium for reconnecting with our emotional experiences of the past. When we recall specific events, it is often accompanied by a re-experiencing of the emotional state at the time. Compared to visual memory, sound can trigger emotions and memories more directly because auditory information is often closely associated with emotional experiences [17]. Previous studies have shown that sound also has a powerful memory-evoking function. For example, Rubin and Wallace [41] pointed out that specific sounds, such as music and speech, can enhance memory for particular events and stimulate emotional associations. Currently, lifelogging focused on visual memory forms has been widely studied and applied [11,45,52]. However, sound, as one of the mediums of memory recording, has been relatively less studied and utilized. These studies on sound-based lifelogging generally rely on a record-and-playback approach [20,33,37], which means that users are unable to reconstruct the sounds.

Relevant research has shown that there are multiple ways to construct memories through sound [47]. Sounds can be associated with specific situations, people, and places, while also being able to evoke broader, non-specific memories of time or events [15]. Thus, for sound memories, in addition to preserving the memory of the moment by recording the sound of the original scene, post-production sound design can also be a creative way of constructing memories. This was confirmed in our subsequent user studies, where we found that through sound design, which typically includes music and sound effects, creators can reconstruct or amplify the sound characteristics of a scene, thereby enhancing the emotional depth and symbolism of the scene in memory. This approach not only preserves the memory of a specific moment, but also provides creators and listeners with richer sensory experiences and emotional associations through sound processing.

In recent years, the popularization of UGC (User-Generated Content) has led to the rise of video creation tools, and people are documenting their lives much more frequently. However, visual media, such as photos and videos, have become the primary means of documentation [5], which has further marginalized sound recording. While many user-generated videos combine sound and images, the sound design aspect is often overlooked and usually serves as background music or voice-over narration rather than as an independent medium for memory recording. Existing sound design software or features are mainly based on audio tracks as a form of interaction, where users need to repeatedly drag and listen to the tracks to get relatively good results, and sound effects, which are typically short in duration and numerous, require even more effort. Moreover, this interaction is not suitable for the operation logic of mobile devices. As a result of these interaction problems, except for some professional video creators, most people do not intentionally explore and create in-depth sound design, especially the use of sound effects.

Based on the above research background, this paper proposes SoundScape, a human-AI co-creation system with a novel interaction that allows users to conveniently create and edit sound memories on mobile devices. The system diagram is shown in Figure 1. SoundScape utilizes an innovative sound memory creation interface that allows users to use images as a creation panel, or a "musical instrument," for sound memory creation. Users simply tap on an object in the image, like pressing a piano key, to add the corresponding sound effect to the timeline without having to repeatedly drag the sound clip along the track. In particular, to assist novice users, SoundScape integrates a conversational agent with AI music generation technology to inspire users and make music creation easier. SoundScape's system design has been validated in a series of user studies. In a controlled experiment, SoundScape provided a better user experience and perceived quality of work in sound memory creation scenarios compared to existing established products.

In summary, this study makes the following contributions:

(1) We propose SoundScape, a human-AI co-creation system for sound memory creation, which integrates artificial intelligence technology to support users in creating sound memories. To our knowledge, SoundScape is the first intelligent sound memory creation system that combines music and sound effect creation.

(2) We propose a novel sound design interaction method, providing a reference for future research on audio-visual integration in human-computer interaction in intelligent art.

(3) We evaluated the user experience and perceived quality of works created with SoundScape through quantitative and qualitative user experiments. The insights and findings from the user studies provide information for research on work perception in human-computer co-creation and emotional expression in intelligent art co-creation.

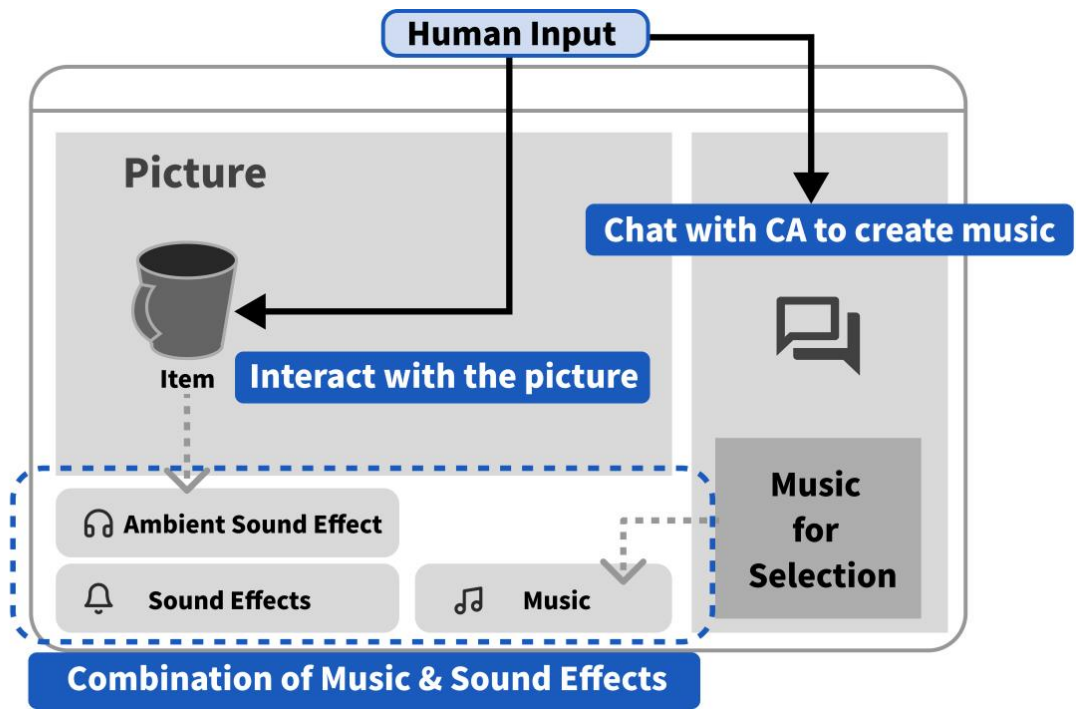

Figure 1: SoundScape System Diagram

## 2 RELATED WORK

### 2.1 Lifelogging and Sound Memories

Lifelogging is the process of documenting the behaviors and activities of an individual's daily life through the use of technology [26]. This recording usually relies on various sensing devices such as wearable cameras [2] and cell phones [3], etc., to capture data related to an individual's sight, sound, movement, heart rate, and more. These recordings can help people recall past events [45], monitor health conditions [7], and even be used in medical and psychological research [38,53].

When users want to search for old memories, they can utilize lifelogging systems for relevant searches. As a result, most research on lifelogging has focused on improving the convenience of the technology—specifically, how optimizing the technology can enhance the effectiveness of memory search [11,18]. However, studies have shown that people are not particularly interested in detailed digital documentation of their lives; instead, they prefer to sift through and reconstruct past life clues [36]. This is because digital recordings often overlook the original purpose of preserving memories, and thus fail to resonate with people on an emotional level. In Hiroshi Ishii's research [25], a tangible user interface was proposed, which integrates physical objects with digital information, creating a novel interaction mode. Consequently, there has been an increasing amount of research focusing on the integration of physical objects and digital information [30,49,52].

Perhaps due to the limited availability of recording devices, these memory evocation studies have largely focused on physical objects and visual media, with relatively little research on sound-based lifelogging. FM radio [37] proposes a new type of digital memento, allowing users to play sounds from past vacations through a retrofitted old radio. Sonic Gems, designed by Oleksik and Brown [33], hides sounds in capsule-shaped containers equipped with RFID tags, and selecting the appropriate capsule triggers the playback of the sound within. Similarly, SoundCapsule [20] transforms the concept of a physical sound capsule into an Android-based mobile application, where users receive their past recorded sound memories in the form of random phone calls. These studies on sound memories generally rely on a record-and-playback approach, and while the mediums differ, the underlying mechanisms are similar. Moreover, in these studies, users are unable to reconstruct the sounds, which means that the presentation of memories doesn't always satisfy people's desire to relive positive moments.

In a study of oral history, the author points out that sound can not only be associated with specific memories, but also evoke a broader range of recollections and even create imaginative spaces that extend beyond the original scene [15]. Therefore, in this study, we aim to move beyond the traditional audio playback format and provide a more creative and open space for the creation of sound memories.

### 2.2 Music and Sound Effects in Sound Design

Sound design plays a vital role in several fields, such as virtual reality [54], gaming [13] and mobile device interaction [27], movies [6], short videos on the web [10], etc. All of these domains have similar needs for sound design, and effective sound design can effectively enhance the experience of the viewer or user, prompting them to be more deeply immersed in the content they are encountering [31]. In particular, music and sound effects are important components of sound design [14], and they play an indispensable role in enhancing the overall atmosphere and immersion [50].

In the film industry, some researchers have successfully enhanced the atmosphere of movies through specific sound design techniques such as synesthesia and metaphor. By carefully orchestrating different sound effects and background music, they have crafted immersive soundscapes that allow audiences to experience emotional depth and heightened tension throughout the film [9] In video games, other researchers have employed various technological tools to further amplify player immersion. For instance, the combination of ambient sounds, character voices, and background music helps players become more integrated into the virtual world, offering a stronger sense of interactivity [13].

A comparative study of different sound design elements found that music creates atmosphere and conveys emotion [51], while sound effects provide specific auditory imagery that breaks through the limitations of video footage, enriches the content, and quickly extends to express the characters' inner thoughts, making the sound narrative more complete [60] sound effects are more effective than music in enhancing immersion [50]. In movie sound design, a fully sound-designed track can quadruple immersion, while a single piece of music has a limited effect [6].

However, on increasingly popular online short video platforms, sound design is primarily focused on music. Although music plays a crucial role in conveying emotions and directing the viewer's attention [61], sound effects are used relatively infrequently. For example, TikTok offers creators numerous video and music filters and effects [46], leading many to prefer popular music as background for their videos, which often results in the potential value of sound effects being overlooked. This trend makes the sound design of short videos feel monotonous, failing to fully showcase the diversity and richness of sound.

Therefore, how to effectively combine music and sound effects to create a richer and multi-layered sound experience has become an important issue to be solved. In this study, we explored new forms of sound design practice by synthesizing these two sound elements, bringing creators more comprehensive creative functions and more immersive effects.

## 2.3 Human-AI Co-creation for Sound

With the rapid development of Artificial Intelligence (AI) technology, human-AI co-creation has seen increasing applications in the field of art.

In the sound effects domain, related research has proposed text-based [16], image-based [22], and video-based [35] sound generation algorithms, offering more efficient methods for sound design. Additionally, further research on interaction and systems has extended these algorithms. The MIMOSA system provides a spatial sound co-creation tool [32] which allows users to easily generate and manipulate spatial sound effects. Hugo et al. [44] also explored human-AI co-creation in sound engineering and the use of computer creativity for audio system design. These studies expand the boundaries of sound design through human-computer interaction and demonstrate the promise of human-AI co-creation for a wide range of applications in audio creation.

In the field of music, the study of algorithms is mainly divided into two major branches: symbolic music generation [19,23,55,57] and audio music generation [1,8,21,42]both of which have seen significant progress. To make AI music technology accessible to a wider audience, researchers have explored the potential of intelligent music interaction systems [10,59,64]. These studies have made complex music generation techniques more actionable through intuitive interactive interfaces that allow users to collaborate with AI, providing more freedom for music creation. In recent years, more AI-driven music creation tools have emerged, allowing users to generate music through tags [65,66], text [62,63], images [67,68], and music [69,70]. The diversity of these generation methods has greatly enriched the application scenarios of human-AI co-creation and provided creators with more efficient tools and support.

AI technology has facilitated the generation of sound materials, but existing systems are mostly focused on the desktop, and there is still a lack of integrated sound design tools that meet the needs of both music creation and sound effect design on mobile devices [4,39]. Current research and products tend to treat music and sound effects separately, requiring users to switch between different tools for different tasks, increasing the complexity of creation. Additionally, the lack of well-designed interactions leads to a steep learning curve for these tools, especially on mobile platforms, resulting in a suboptimal user experience [28]. Therefore, future research should focus more on music and sound co-creation in mobile environments, providing creators with more convenient and user-friendly tools through integrated interaction design and innovative algorithms, ultimately enhancing the creative experience.

## 3 FORMATIVE STUDIES

### 3.1 Focus Group

*3.1.1Method.*

After identifying the research theme of "Sound Memory Creation" through a literature review, we conducted a focus group workshop with the researchers [34]. The aim of the focus group was to further understand the needs and potential challenges in the process of audio memory creation, as well as to brainstorm and hold conceptual discussions based on these needs, seeking guidance for the design of the user interface of the envisioned interactive system for audio memory creation. A total of eight researchers (three men and five women) participated in the workshop. They are engaged in human-computer interaction, design, music audio algorithms, music education, and related fields, possessing extensive knowledge in sound creation and human-computer interaction. This enabled them to offer valuable insights from various professional perspectives.

During the workshop, the facilitator provided a general introduction to the project, including the research background, previous findings, and the target user group (users interested in recording their daily lives but with limited sound design skills). The discussion then focused on three major themes: (1) the importance of sound in life recording creation; (2) the problems and needs of current apps used in life recording, particularly during the sound design phase; and (3) brainstorming and further defining potential interaction strategies, functions, and user interfaces for sound memory creation in the context of life recording. The session lasted for two hours, with the first author taking notes and further summarizing the discussion afterward.

*3.1.2 Results.*

When discussing the role of sound in lifelogging creation, participants fully shared their feelings and evaluations of sound as a creative medium. First of all, the participants generally agreed that their primary goal when creating and documenting was to capture those good memories, regardless of whether they would later share them on social platforms. Therefore, being able to better capture and present their memories was what everyone sought in the creative process. The first author further summarized this part of the discussion and identified the following roles of sound in lifelogging: (1) Memory evocation: compared to purely visual information, images with sound can help evoke memories more quickly and with greater completeness; (2) Emotional resonance: multimodal information can enhance the immersion and atmosphere of photos, making emotional expression more vivid and eliciting greater empathy from viewers; (3) Enhanced expressiveness: compared to static photos, sound as a time-based medium brings more experiential dimensions and adds dynamic continuity to static works, making sound-enhanced works more attractive to viewers than static images alone.

Next, participants further discussed the current issues they encountered when using lifelogging and creative software. They mentioned that photos remain their main form of documentation in daily life because taking pictures with a smartphone is a simple process that doesn't require much time or effort. However, this efficiency also results in a loss of certain information (e.g., time flow, sound, etc.), which requires post-editing. Current mobile editing tools, however, are mostly adapted from the interaction logic of video editing software, making them inconvenient to use. While automated templates can generate content with a single click, they limit the user's creative freedom. Additionally, with features like iPhone's Live Photos, participants pointed out that these often include unwanted sounds or noise. Moreover, since they don't typically use high-quality recording equipment during shooting, the captured sound tends to be of poor quality. Therefore, for a Sound Memory Creation System, participants hoped it would offer a simple interaction method that still allows for creative autonomy. It would also be ideal if the system could provide high-quality sound effects and music.

During the brainstorming session, many ideas for the functionality of a sound memory creation system were proposed, including some interesting and valuable ones. First of all, based on the conclusions of the previous discussion, everyone felt that utilizing photos for sound memory creation is a better form of creation. On the one hand, photos are one of the most popular ways to document life, and working with photos ensures users have a reliable source of creative materials. On the other hand, sound is relatively abstract and subjective, while photos are a visible and concrete medium, so using photos as a foundation for creation can provide users with a more intuitive and controllable starting point.

One participant suggested framing an object within an image and then assigning a sound effect by clicking on it, an interesting idea that was well-received by the group. Additionally, several participants expressed the desire to use AI capabilities to improve the effectiveness and efficiency of creation. Some proposed that AI music generation technology could be used to create music that matches the atmosphere of the scene depicted in the image. Another participant suggested integrating the currently popular LLM technology to enable users to describe the sound memories they want to create to a conversational agent (CA), which would then generate the corresponding audio content.

Of course, there were also some differing opinions during the brainstorming. Some participants thought that sound memory creation should be based on a precise timeline, allowing users to fine-tune various sound elements. However, another participant argued that sound effects should be edited more simply, stating that "photographs don't inherently carry temporal information, so does the creator care if a sound element appears at a specific second? I don't think so. Timeline-based editing is also more complex by nature. I think it's possible to display all sound effects for users to easily review, and then allow them to set looping patterns or the order of the sounds." This novel idea also provided a new perspective on how sound memory creation could be approached.

### 3.2 Interaction Scenarios

*3.2.1 Method.*

Based on the ideas presented at the focus group discussion, the researchers defined potential interaction scenarios for sound memory creation and invited users to evaluate them. We proposed two possible approaches for both the music generation phase and the sound editing phase, as shown in Figures 2 and 3, respectively.

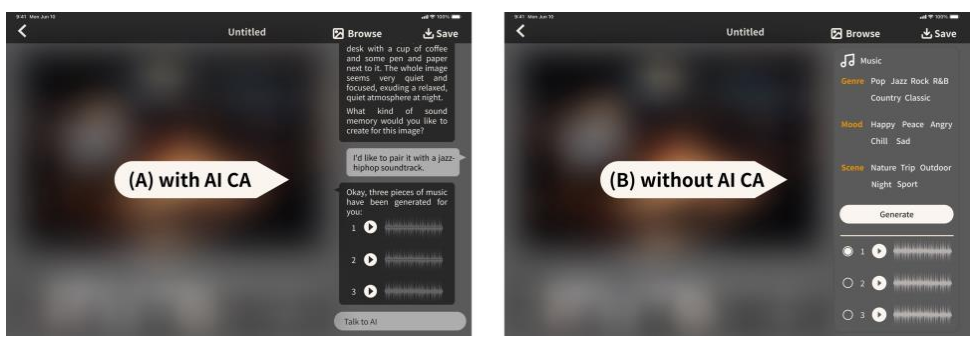

Figure 2: Two approaches for music generation. (A) With AI CA: the user provides a natural language description to the AI CA, and the system intelligently generates music; (B) Without AI CA: the user selects labels from three dimensions—scene, style, and emotion—and the system intelligently generates music.

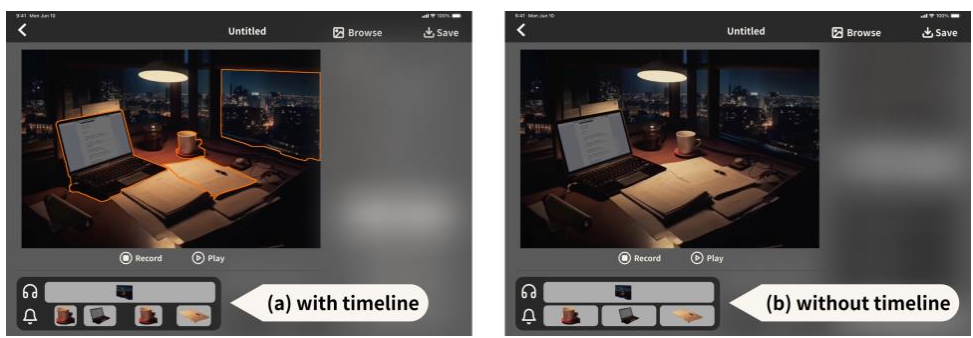

Figure 3: Two approaches for sound effect editing. (a) With timeline: sound effects are displayed on a precise timeline, and the user independently adjusts the timing of the sound effects; (b) Without timeline: sound effects play in the order arranged by the user, without a precise timeline.

We distributed the interaction scenarios questionnaire on the public data platform Credamo [71], and users who did not have the habit of recording their lives (e.g., taking photos, videos, etc.) were excluded from participation. We received 64 questionnaires, and after excluding responses that were too short, careless, or did not meet the target demographic criteria, we ended up with 60 valid questionnaires (23 males and 37 females). Among them, 40 users were under 30 years old, 17 were between 31 and 50 years old, and 3 were over 50.

The questionnaire consisted of single-choice questions, where participants selected the interaction scheme they preferred, and open-ended questions, where they described their perceptions of each interaction scenario. In the study, we presented the different interaction schemes for the two stages of sound memory creation via videos, providing descriptive cues for the respective workflows in each video. In order to reduce the risk of carryover effects, the two scenarios in each stage were shown in a randomized order, ensuring that all four possible order combinations were assigned to 15 participants each. Finally, we asked participants to share their opinions and suggestions regarding the overall functionality of the system. For the questionnaire data processing, we analyzed the single-choice questions by frequency analysis, while the free-text responses were processed using keyword coding.

*3.2.2 Results.*

The main categories of opinions are elaborated in the table below.

Table 1: Advantages and disadvantages of the scenario

| Scenarios | | Advantages | Disadvantages |
| --- | --- | --- | --- |
| Music Generation Phase | (A) With AI CA | Convenient/Easy | Requirements for prompt ability |
| | | Personalization/Creativity | Complexity of the description |
| | | High degree of freedom/flexibility | Limiting creativity |
| | (B) Without AI CA | Intuitive/efficient | Limited options |
| | | Simple and easy to operate | Lack of individualized regulation |
| | | No musical skills required | Music knowledge required |
| Sound Editing Phase | (a) With Timeline | Higher precision/better results | Requires repeated adjustments |
| | | High degree of flexibility/freedom | take a period of (x amount of time) |
| | (b) Without timeline | Convenient/Easy | Little room for regulation |
| | | Reduced user actions | Poor artwork |

In the music generation phase, scenario (A) (with AI CA) was favored by most participants (47/60). This is because it allows users to freely express their ideas during sound memory creation, and the ability to continuously interact with the CA fosters greater inspiration, ultimately generating personalized works that align with the user's creative intent. One participant stated, "The scheme allows me to more intuitively express my personal emotions and needs through natural conversation. The AI agent can understand and generate music that better matches the user's intentions, enhancing the interactivity and personalization of the creative process." Additionally, communicating with the CA in natural language, as one participant described, gives a greater sense of companionship and aligns with the concept of emotional design. She said, "With the AI agent, it feels like having company, less lonely. Editing videos with music feels more relaxed and can spark inspiration. Without the AI agent, the interface is simpler but dull, and when I don't know what kind of music to use, it feels frustrating and uninspiring."

In the sound editing phase, scenario (a) (with timeline) was supported by more users (48/60). The sound editing method with a timeline allows for more precise effects, such as better synchronization with the background music, and provides users with greater creative freedom. Users can add sound effects at the right moment according to the overall emotional atmosphere, even though this approach may take more time to complete. One participant commented in the questionnaire, "This scenario provides more flexibility and precision, allowing me to accurately place sound effects on the track according to the tempo and emotional changes of the music, resulting in sound memory pieces with more depth and dynamic effects.

While this method might require more time for fine-tuning, it better meets the needs of professional users and those aiming for more refined creations." Furthermore, option (a) requires users to click on objects in the image to add corresponding sound effects, making the interaction more engaging, with one participant describing it as "very dynamic."

Overall, participants gave positive feedback on our sound memory creation interactive scenario, which provided users with an intelligent tool for creating ambient, immersive, and emotive sound memories based on images. One participant said, "This really hit me straight in the heart. I think the overall rhythm and atmosphere are great, I love it."

### 3.3 Summary of Findings

Summarizing the user study described in the previous two sections, we draw the following conclusions:

- Audiovisual content provides a better sense of atmosphere and helps users memorize and evoke emotions more quickly. Users are more receptive to and enjoy these types of media formats.
- Photos are one of the most common ways for users to document their lives, and they generally prefer using photos as the foundation for secondary editing to create life records.
- The sound in videos or audio photos often contains noise and unwanted sounds, prompting users to perform secondary sound editing.
- The main challenges users face when adding sound to photos include complicated editing processes, a lack of personalization in their work, and a shortage of high-quality sound materials.
- Users believe that AI technology can assist in the intelligent generation of sound materials; additionally, conversational agents can offer creative support, such as providing inspiration, while maintaining creative freedom.
- While ensuring ease of use, users hope that the creation system can offer more detailed creative features to achieve better results, rather than simply focusing on making the process easier.

## 4 SYSTEM DESIGN

### 4.1 Design Process

We followed the concept of "user-centered design" throughout the design process, and a series of user studies inspired and informed the system design. In addition to the eight researchers in the focus group, a total of 74 general creators participated in the design process, including 60 participants in the interaction scenarios and 14 participants in the system evaluation. None of the general creators were related to the project researchers. After identifying the importance of sound memory creation, the project team researchers first organized a focus group discussion on the potential problems with existing creation software in sound memory creation scenarios and a brainstorming session to envision the functionality, interface, and interaction of the sound memory creation system. The discussion results indicated that track-based editing on mobile devices is very challenging for creators. The highly templated approach to creation also limited creators' creative freedom. Additionally, higher quality music and sound effects sources were also desired. Based on these findings and the ideas from the brainstorming session, we proposed two interaction scenarios for both the music creation phase and the sound editing phase and asked a group of general creators to choose between them. The results showed that participants preferred CA-assisted creation and the use of a precise timeline to achieve better results for their works. Finally, the system passed the evaluation by another group of general creators. The final system UI design and functional structure are shown in Figures 4 and 5. A system demo video will be provided in the attached materials.

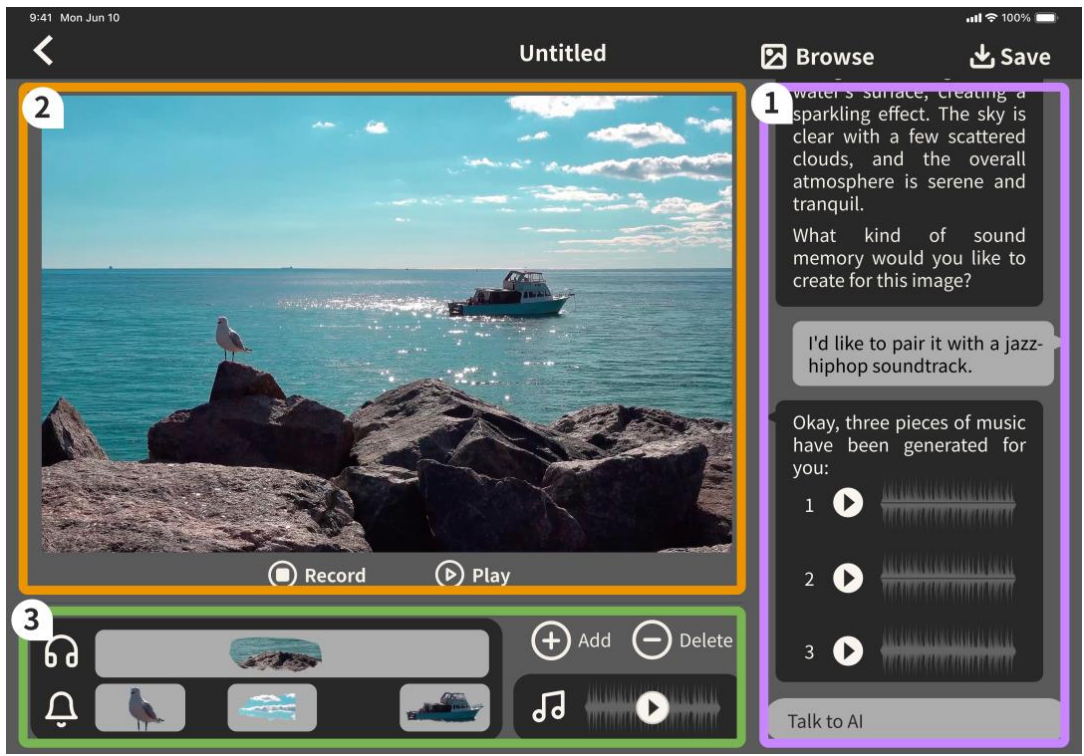

Figure 4: SoundScape system UI design. (1) Conversational Agent Area, (2) Interaction Stage Area, (3) Sound Clip Area.

### 4.2 Workflow and Development

The SoundScape system consists of an iOS application based on SwiftUI, which allows for the rapid construction of user interfaces using declarative syntax and enhances the performance and interactive experience of the application through deep integration with system features. The system is divided into three main areas:

1. Conversational Agent Area: Users can interact with the CA in this area, where the CA will interpret the image uploaded by users and assist them in idea generation and background music creation. The built-in dialog agent module supports scene recognition and music generation through the ChatGPT API [72] and Suno-based API [73] respectively. Users first select or take a photo, and the dialog agent recognizes the scene in the image, provides a description, and asks, "What kind of sound memory do you want to create?" Based on user feedback, the CA generates music and offers three options for users to choose from. If the user is unsatisfied with the generated music, further descriptions can be provided to adjust the results iteratively.

2. Interaction Stage Area: This area displays the user's created sound memories, with object recognition and sound effect addition modules integrated for user interaction. The object recognition module uses the open-source YOLOv3 [40] scheme and employs CoreML-based offline models to quickly recognize objects. The sound effect addition module contains a pre-built database of object-sound pairs, sourced from recorded materials or open-source platforms. Users can select objects in the photo to add sound effects by drawing a box around them, after which the system identifies the object type and retrieves matching sound effects for users to choose from.

3. Sound Clip Area: All sound assets used in the sound memory creation will be displayed in this area. It consists of three separate music players for background music, ambient sound effects, and regular sound effects, and includes a timeline for real-time recording of sound effect trigger signals. When the user starts recording, the system automatically plays the pre-set background music and ambient sounds (if set). By clicking on an object in the Stage area, the system instantly marks the corresponding sound trigger on the timeline and integrates it into the audio track. Once the recording is complete, users can preview the result by clicking the Play button.

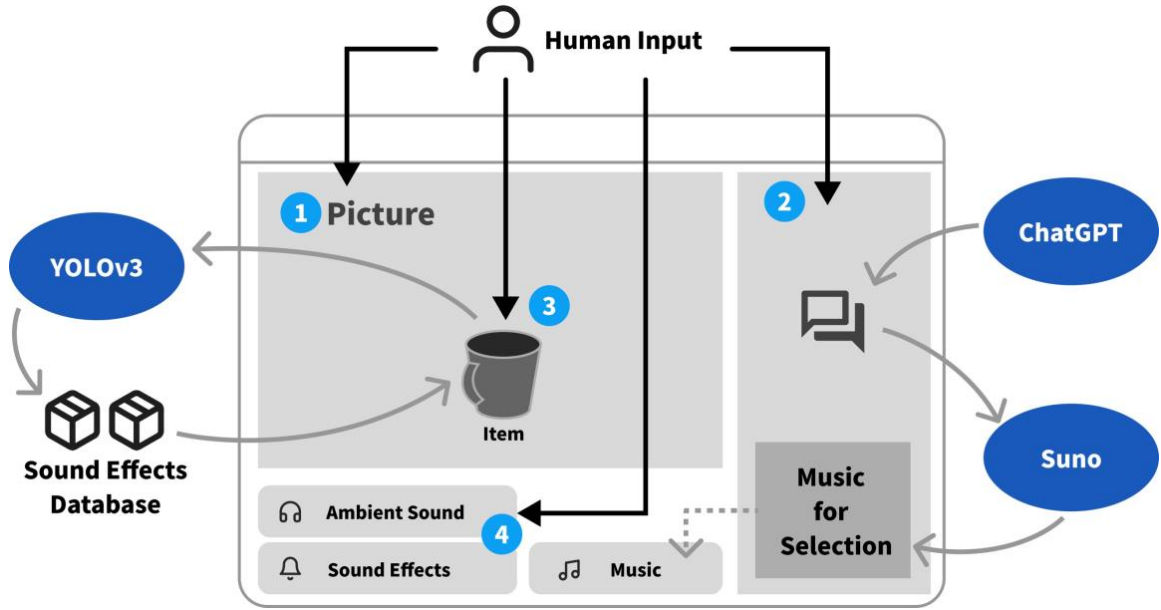

Figure 5: Functional structure of the SoundScape system: (1) capturing or importing images; (2) interacting with the CA and generating background music; (3) adding sound effects by boxing objects in the stage area; (4) recording the sound memory creation.

## 5 EVALUATION

### 5.1 Method

The system prototype evaluation of SoundScape took the form of a comparative experiment with the following two objectives: (1) to validate the usability and user experience of SoundScape; and (2) to compare SoundScape with existing authoring software to verify whether, in the scenario of sound memory creation, SoundScape's usability and user experience are significantly better than those of existing authoring software. In this experiment, the researchers chose the CapCut app as a comparison for SoundScape. Since SoundScape is designed to address issues in sound scene construction and system interaction, the comparison software for the experiment needed to meet the following conditions: (a) it can handle both soundtrack and sound effect design; (b) it has an interactive interface. Considering these two points, we decided to use video editing software for comparison, as such software typically has mature interaction designs and complete functionality, which can meet the needs of sound memory creation scenarios. After conducting research, we chose the CapCut app as the comparison for SoundScape. According to an industry research report, CapCut ranked first in usage in the Top 50 AI APP Products in China in July 2024 [74], holding a significant share of the Chinese user market. Therefore, the CapCut app is highly representative.

The system prototype evaluation took about 30 minutes per person, and participants were required to complete all testing sessions on an iPad. The experimental procedure was as follows: (1) Participants signed a consent form and read the experimental instructions (containing the background, preparation, and procedure of the experiment). (2) Then they watched a demonstration video of the SoundScape system. (3) Creation Task 1: use the SoundScape system to create a sound memory based on an image. (4) Creation Task 2: use the CapCut app to create a sound memory with similar content based on the same image from Task 1. (5) After completing the two tasks, we conducted a semi-structured interview to gather participants' feedback on the functions and interactions of the SoundScape system, as well as the problems they encountered. (6) Finally, participants filled out a questionnaire to assess the usability and user experience of both systems. Participants received a cash reward of 50 RMB (about 7 USD).

In the usability and user experience section of the questionnaire, we used the short version of the User Experience Questionnaire (UEQ-S) [43]. The UEQ-S contains eight items, each consisting of a pair of antonyms, with the first four items representing the pragmatic quality scale and the last four representing the hedonic quality scale. In addition to the eight items in the UEQ-S, we added an extra item designed to assess the perceived quality of the output from the two interactive systems in a sound memory creation scenario. All items were rated on a 7-point Likert scale. To facilitate uniform calculation, all items were scored on a scale from -3 to 3, so a positive score indicated a positive user evaluation of the system on that dimension. The questionnaire, along with the quantitative data from the evaluation session, will be included as supplementary material.

Although previous studies have validated the items in the UEQ-S, we conducted a confirmatory factor analysis (CFA) of the variables to ensure their validity. The results showed that the factor loadings of all items were greater than 0.6 and were significant, which meets the accepted standards, indicating that the items demonstrate strong measurement relationships [48]. Cronbach's alphas for each variable were greater than 0.8 [12] and the average variance extracted (AVE) was greater than 0.6 [24], showing that the items in the UEQ-S also exhibited good internal consistency and convergent validity.

### 5.2 Results

We posted a call for user research on our social media platforms and gathered a total of 14 users (7 females and 7 males) to participate in this evaluation of the system prototype. According to Macefield's review [29], for comparative studies, 8-25 participants per group is a reasonable range to consider, with 10-12 participants being a good baseline. Thus, the sample size for this evaluation experiment is valid. The participants were predominantly young, with an average age of 25.7 years (median 25.5, range 22-30 years). Among these participants, 2 were professional music creators, 6 had some experience experimenting with music creation but were not professional, and 6 had no experience in music creation at all. In terms of video creation, 4 had extensive experience, 7 had some experience, 2 had limited experience, and 1 had almost no video creation skills. The vast majority of participants had the habit of taking photos or videos to document their lives (3 had done so almost every day over the past month, 6 had documented 1 to 3 times per week, 2 had done so about once a month, and 3 had not recently documented their lives).

### 5.2.1 Quantitative Analysis.

Table 2 shows the results of the paired t-test analysis of SoundScape and CapCut. From the results in the table, it is clear that SoundScape significantly outperforms CapCut in all scores on the dimensions of pragmatic quality, hedonic quality, and perceived quality of the work.

Table 2: Results of paired t-test analysis of (a) SoundScape and (b) CapCut

| Indicator | Mean ± Standard Deviation | | Mean Difference | t | p |
|---|---|---|---|---|---|
| | a | b | | | |
| a PQ 1 pairing b PQ 1 | 2.14±0.66 | 0.71±0.99 | 1.43 | 4.372 | 0.001** |
| a PQ 2 pairing b PQ 2 | 1.21±0.97 | -0.50±1.61 | 1.71 | 3.067 | 0.009** |
| a PQ 3 pairing b PQ 3 | 2.00±0.78 | -0.07±1.07 | 2.07 | 5.597 | 0.000** |
| a PQ 4 pairing b PQ 4 | 1.21±0.97 | -0.07±1.33 | 1.29 | 2.432 | 0.030* |
| a PQ pairing b PQ | 1.64±0.73 | 0.02±1.04 | 1.63 | 4.064 | 0.001** |
| a HQ 1 pairing b HQ 1 | 2.00±0.78 | 0.00±1.11 | 2.00 | 5.292 | 0.000** |
| a HQ 2 pairing b HQ 2 | 2.21±0.80 | -0.07±0.92 | 2.29 | 5.95 | 0.000** |
| a HQ 3 pairing b HQ 3 | 2.29±0.73 | -1.07±1.44 | 3.36 | 7.632 | 0.000** |
| a HQ 4 pairing b HQ 4 | 2.29±0.73 | -0.79±1.37 | 3.07 | 7.704 | 0.000** |
| a HQ pairing b HQ | 2.20±0.68 | -0.48±1.14 | 2.68 | 7.184 | 0.000** |
| a PQW pairing b PQW | 2.21±1.05 | -0.07±1.44 | 2.29 | 4.824 | 0.000** |

[a] * $p<0.05$ ** $p<0.01$

[b] PQ stands for Pragmatic Quality, HQ stands for Hedonic Quality, and PQW stands for Perceived Quality of the Work

Figures 6 through 8 show the frequency histograms of SoundScape's four pragmatic quality scores, four hedonic quality scores, and the perceived quality of the work scores. It can be seen that SoundScape's scores are all 0 and above, proving that participants hold a more positive opinion of both SoundScape's user experience and the perceived quality of the work.

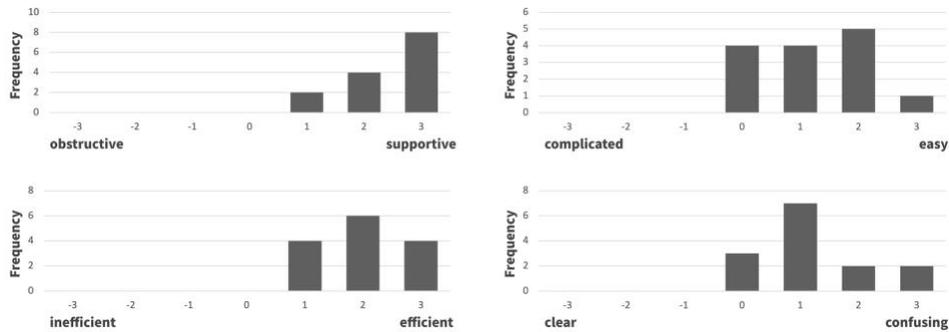

Figure 6: User ratings of SoundScape's four pragmatic qualities, ranging from -3 to 3, with the meanings at each end of the scale indicated in the figure.

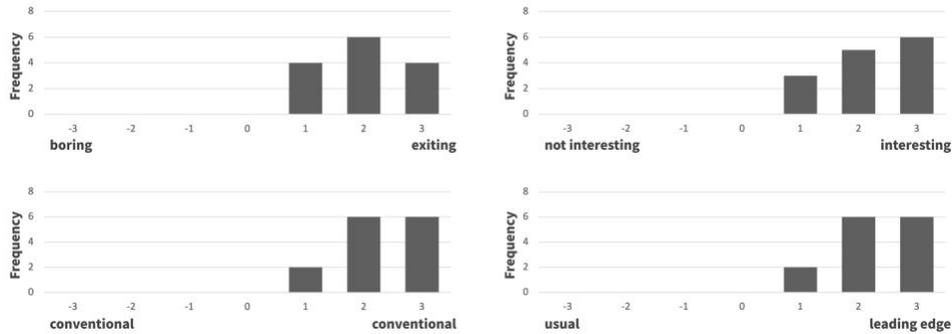

Figure 7: User ratings of SoundScape's four hedonic qualities, ranging from -3 to 3, with the meanings of the two ends of the scale indicated in the figure.

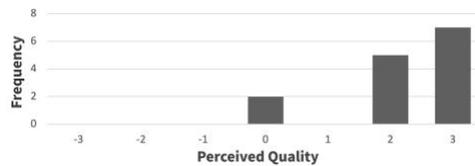

Figure 8: User ratings of perceived quality of SoundScape works, ranging from -3 = not good at all, to 3 = very good

*5.2.2 Qualitative Analysis.*

**Creation process.** In the creation task, we asked the participants to choose an image for their creation, so the content of each piece was different. Observing the creative process, we found that most participants would mentally construct a scene based on the image they chose, feel the atmosphere of the scene, and then think about what sounds would fit in. Since most of the photos chosen by the participants were not recently taken, they wouldn't clearly remember what sounds were present at the time the photos were taken. Therefore, we observed that some participants used SoundScape to get inspiration by talking to a conversational agent. When describing the background music, most participants focused on emotions, style, etc., while some described the scene in the picture more concretely, allowing SoundScape to generate music that matched the image.

**Artwork.** The number of sound effects that participants added to their works ranged from two to five. Although ambient sounds were not required, almost all participants added sounds like wind and rustling leaves to their works. They stated that adding ambient sounds effectively enhanced the atmosphere of their work and brought a more realistic sense of immersion to the sound memory project. "The background sound effects make me feel like I'm in that scene. Without a background sound constantly ringing in my ears, it feels too quiet and not realistic enough."

Although in most cases the sound effects recommended by SoundScape met the participants' needs, they wished for more options and the ability to browse the sound effects library directly. Additionally, participants expressed a desire for more editing capabilities for music and sound effects, as there were times when the volume of a sound element was too loud or too quiet, which affected the overall presentation of the sound memory project.

**Interview Evaluation.** Based on the interview results, the most common feedback from participants regarding SoundScape was that it is "interesting and convenient." Although CapCut is one of the most commonly used mobile video editing apps today, it still presents many interaction challenges. SoundScape, on the other hand, allows users to directly interact with images in the stage, offering more innovation and playability. It also adapts well to mobile touch screen interaction, avoiding the need for precise timeline dragging, making it more user-friendly than traditional software interactions. However, many participants mentioned that SoundScape lacked sufficient guidance during use, and that the interface's functions were not clearly organized. Even after watching the demo video, some users still felt confused during their first attempt, though this confusion tended to diminish after a few tries.

Additionally, the form of human-AI co-creation makes SoundScape's outputs more tailored to individual needs. Participants noted that the soundtracks generated by SoundScape were mostly unique and aligned with their expectations, and even if they weren't satisfied with the initial result, they were able to generate a satisfying piece of music after two or three iterations. In contrast, while apps like CapCut boast a powerful and extensive library of music and sound effects, they often recommend currently trending music, which may not match the user's specific needs. As a result, users often find themselves spending a long time "searching and previewing," often ending up with nothing. One participant commented, "The conversation with the conversational agent made me feel like the system understood what I wanted, even though I'm not a professional in music. This co-creation process also allowed me to create something personal, which aligns with the concept of sound memory, as memory is a deeply personal experience."

### 5.3 Summary of Findings

Through quantitative evaluation, we find that SoundScape outperforms CapCut in terms of user experience and perceived quality of the work, demonstrating that SoundScape offers a significant improvement over existing audiovisual creation products in sound memory creation scenarios.

The findings of the qualitative study also corroborate the quantitative assessment results. While there is still room for improvement, SoundScape generally meets the creative needs of its users. It offers strong creative support for creators, and compared to existing audiovisual creation software, SoundScape provides a more personalized interactive experience.

## 6 DISCUSSION

In this study, we designed and developed SoundScape, a human-AI co-creation system with a novel interaction approach that allows users to easily create sound memories on mobile devices. We adopted a user-centered design approach and developed the system based on a series of user studies, identifying the most suitable interaction design from four scenarios. SoundScape combines AI music generation technology with an LLM-based conversational agent, which assists novice users without a background in sound design in their creations. Notably, we de-emphasized the traditional sound track editing model and proposed a more innovative interaction method. By transforming the stage into an authoring interface, users can add sound effects to the timeline by simply tapping on objects in the screen, reducing the difficulty of operation while enhancing real-time interactivity. Empirical studies show that SoundScape meets users' needs in sound memory creation scenarios and outperforms existing tools in terms of user experience and perceived quality of the work.

**Perception of human-AI co-creation in special scenarios.** Our user study shows that for personalized private creation, Human-AI co-creation can provide a better creative experience and a higher perceived quality of works—users do not want their works to be identical, and the unique works co-created with the AI precisely meet their psychological needs. In our experiments, although the control system had a much more mature and extensive material library, participants did not view this as an advantage. Instead, they found the process of finding suitable material time-consuming, and felt they were merely integrating materials rather than creating something original. This is an interesting finding because most current authoring tools focus on improving efficiency and effectiveness, rather than the uniqueness and privacy of the work. In certain scenarios, the perceived quality of a work may be influenced by self-efficacy and the uniqueness of the creation.

**Sound Memory and Emotional Links.** Sound plays an important role in memory, and specific sounds can serve as memory cues that evoke emotions and memories [15], helping people recall the past. In the special context of sound memory creation, users, as both the subject of memory and the creator, pay closer attention to their role in the creation and the feelings the creative process brings them. In one experiment, a participant chose a childhood photo to create with. When discussing her feelings about the creation process, she said, "Actually, I can't remember the sounds from when I took this photo, but after adding music and sound effects with my imagination, I felt like I was back in that innocent and carefree time." This shows that the creation of sound memories deepens emotional engagement, making their emotional world richer and sometimes allowing them to gain new insights. In this study, we provided the public with a Human-AI co-creation tool for sound memories, which not only reduces the difficulty of sound memory creation but also helps users build and strengthen their emotional connections to the work's scene. Intelligent artworks are often criticized for lacking soul and emotion, and our findings on the creation process and emotional links may serve as a reference for future research.

## 6.1 Potential Roles

In the user test interviews, one participant reported feeling emotionally relaxed while composing with SoundScape, and that the novelty and fun of the experience made her feel "like she was playing a less intense musical game." Additionally, for some of the longer sound-memory pieces, the creator mentioned that they were great as background white noise for study sessions. If SoundScape could intelligently extend and loop his compositions, he would gladly use it as background music while working in the office to improve his concentration and productivity. Therefore, the researchers believe that beyond creating sound memories, SoundScape could serve additional functional roles to help users relieve stress and enhance efficiency.

Another interesting idea caught the researchers' attention. One participant said, "The process of creating with SoundScape actually turns the image into sound, and it might be able to do something for the blind. ...... They can't see the image, but we can use SoundScape to let them hear it." This point greatly inspired the researchers, as sound can enhance the media experience in situations where visual elements are limited [50] . Therefore, describing the content of a picture for the visually impaired through sound and enabling them to perceive more high-level information about the picture might be one of the future directions for SoundScape, which could also bring more meaning to SoundScape.

## 6.2 Limitations and Future Work

First of all, since SoundScape's sound-adding mechanism relies on images, creators may encounter difficulties when they want to add sound effects for objects that don't appear in the image. One possible solution is to let users add off-screen sound effects by interacting with a conversational agent. In the future, we will also explore how to enable users and AI to reason through the creative process, leading to richer and more imaginative creations.

Second, SoundScape's music generation functionality relies on Suno, so the black-box nature of the AI model inevitably introduces a degree of randomness [56], which means that SoundScape's generated results may not always fully meet the creators' requirements, even when the results are highly refined. However, due to SoundScape's modular design, the music generation model can be updated and replaced as the technology evolves, improving the controllability of the generation. Additionally, studies on iterative interactive composition [58] have introduced more interactive AI music composition models, providing valuable insights for future prototyping studies.

Third, the system evaluation session was not conducted in a randomized order, and all participants used the SoundScape system before the CapCut system, which may have introduced a carryover effect that could have influenced the evaluation of both systems. The researchers' primary concern was to ensure that participants would use elements similar to those in SoundScape when using CapCut, as CapCut also contains many features unrelated to the sound memory creation scenario. If participants used additional features, it could have reduced the relevance of the experimental data. In the future, consideration could be given to conducting a between-group experiment and limiting the system features of the control group to ensure that the feature sets of the two systems are similar, thereby reducing any subjective carryover influence on the participants.

Fourth, the participants in the user study for the prototype system evaluation were all young individuals from China with good educational backgrounds. However, the usage habits of younger groups with mobile devices may differ from those of other demographic groups. Additionally, regional cultures may also influence creative habits and preferences; therefore, the findings in this study may introduce bias and may not necessarily apply to other groups. Future work may involve assessing populations beyond young Chinese individuals for further validation.

## 7 CONCLUSION

This paper presents a user-friendly human-AI co-creation system, called SoundScape, which allows users to create personalized sound memories on mobile devices using images. SoundScape employs an innovative interface that lets users use images as creation panels for sound memory composition. To assist novice users, SoundScape combines a conversational agent with AI-driven music generation technology, helping users quickly gain inspiration and compose music. This work was developed using a user-centered design approach, with several user studies serving as the basis for system design decisions. The usability of SoundScape was validated through case studies, and compared to existing mature products, SoundScape offers a superior user experience and perceived quality of works in sound memory creation scenarios. This study represents a notable exploration in the field of intelligent art and human-computer co-creation, offering novel insights. We hope that this work will inspire more research on emotionalization in intelligent art within specific application scenarios, and contribute to the popularization of human-computer art co-creation.